\documentclass[twocolumn]{article}
\usepackage{amsmath,epsfig,psfig,url,graphicx,usenix}
\usepackage[T1]{fontenc}
\usepackage{ae,aecompl}

\renewcommand{\footnotesize}{\small}

\makeatletter
{\catcode`\/\active\catcode`\_\active\catcode`\.\active
\gdef\URLslash{\futurelet\next\@@URLslash}%
\gdef\@@URLslash{\ifx\next\URLslash\char`\/\else\slash\fi}%
\gdef\URLdot{\char`\.\penalty\exhyphenpenalty}%
\gdef\URLprepare{\catcode`\/\active\catcode`\_\active\catcode`\.\active
        \let/\URLslash\let.\URLdot\def~{\char`\~}\def_{\char`\_}}}%
\def\URL{\bgroup\URLprepare\realURL}%
\def\realURL#1{\tt #1\egroup}%

\def\Section {\S}

\newsavebox{\sbrack}
\newsavebox{\mbrack}
\newsavebox{\lbrck}
\newsavebox{\mlbrck}

\title{\sffamily\fontsize{18}{18}
  \textbf{Tycoon: an Implementation of a Distributed, \\ Market-based Resource Allocation System}}

\newcount\aucount
\newcount\originalaucount
\newdimen\auwidth
\newdimen\auskip
\newcount\auskipcount
\newdimen\auskip
\global\auskip=1pc
\newdimen\allauboxes
\allauboxes=\auwidth
\newtoks\addauthors
\newcount\addauflag
\global\addauflag=0 

\gdef\numberofauthors#1{\global\aucount=#1
\ifnum\aucount>4\global\originalaucount=\aucount \global\aucount=4\fi 
\global\auskipcount=\aucount\global\advance\auskipcount by 1
\global\multiply\auskipcount by 2
\global\multiply\auskip by \auskipcount
\global\advance\auwidth by -\auskip
\global\divide\auwidth by \aucount}

\newfont{\affname}{phvr at 10pt}
\newfont{\eaddrfnt}{phvr at 8pt}

\author{
\small Kevin~Lai~~~~~~Lars~Rasmusson~~~~~~Eytan~Adar~~~~~~Stephen~Sorkin~~~~~~Li~Zhang~~~~~~Bernardo~A.~Huberman\\
{\small \{klai, lars.rasmusson, eytan.adar, stephen.sorkin, l.zhang, bernardo.huberman\}@hp.com}\\\\
HP Labs Palo Alto
}

\global\def\@maketitle{%
  \newpage
  \begin{center}%
  \let \footnote \thanks
\expandafter\ifx\csname acmdescription\endcsname\relax
  \null
\else
  {\setbox0\hbox{\vbox{%
\begin{flushright}%
  \begin{tabular}[t]{r@{}}%
    \acmdescription
    \\ \noalign{\vskip0.25in}%
  \end{tabular}%
\end{flushright}%
\null}}\ht0=0pt\dp0=0pt\box0}%
\fi
    \vskip -1.2em%
    {\LARGE \@title \par}%
    \vskip 2em%
    {\large
      \lineskip .5em%
      \begin{tabular}[t]{c}%
        \@author
      \end{tabular}\par}%
    \vskip 1em%
    {\large \@date}%
  \end{center}%
  \par
  \vskip 1em}
\newcommand{\beq}{\begin{equation}}
\newcommand{\eeq}{\end{equation}}
\newcommand{\benq}{\begin{eqnarray}}
\newcommand{\eenq}{\end{eqnarray}}

\def\ged{\hbox{${\vcenter{\vbox{
        \hrule height 0.4pt\hbox{\vrule width 0.4pt height 6pt
        \kern5pt\vrule width 0.4pt}\hrule height 0.4pt}}}$}}

\def\compactify{\itemsep=0pt \topsep=0pt \partopsep=0pt \parsep=0pt}
 \let\latexusecounter=\usecounter

\begin{document}

\maketitle

\thispagestyle{empty}

\begin{sloppypar}

%
%

\begin{abstract}

Distributed clusters like the Grid and PlanetLab enable the same
statistical multiplexing efficiency gains for computing as the
Internet provides for networking. One major challenge is allocating
resources in an economically efficient and low-latency way.  A common
solution is proportional share, where users each get resources in
proportion to their pre-defined weight. However, this does not allow
users to differentiate the value of their jobs. This leads to economic
inefficiency. In contrast, systems that require reservations impose a
high latency (typically minutes to hours) to acquire resources.

We present \emph{Tycoon}, a market based distributed resource
allocation system based on proportional share. The key advantages of
Tycoon are that it allows users to differentiate the value of their
jobs, its resource acquisition latency is limited only by
communication delays, and it imposes no manual bidding overhead on
users. We present experimental results using a prototype
implementation of our design.

\end{abstract}

\section{Introduction}
\label{sec:introduction}

A key advantage of distributed systems like the Grid \cite{foster1997}
and PlanetLab \cite{planetlab2003} is their ability to pool together
shared computational resources. This allows increased throughput
because of statistical multiplexing and the bursty utilization pattern
of typical users. Sharing nodes that are dispersed in the network
allows lower delay because applications can store data close to
users. Finally, sharing allows greater reliability because of
redundancy in hosts and network connections.

The key issue for shared resources is allocation. One solution is to
add more capacity. If resources are already optimally allocated, then
this is the only solution, albeit a costly one. In all other cases,
allocation and additional capacity are complementary. In addition, in
peer-to-peer systems where organizations both consume and provide
resources (e.g., PlanetLab), careful allocation can effectively
increase capacity by providing assurances to reluctant organizations
that contributions will be returned in kind.

However, resource allocation remains a difficult problem. The key
challenges for resource allocation in distributed systems are:
\emph{strategic} users who act in their own interests, a rapidly
changing and unpredictable demand, and hundreds or thousands of
unreliable hosts that are physically and administratively distributed.

Our approach is to incorporate an economic \emph{mechanism}
\cite{hurwicz1973} (e.g., an auction) into the resource allocation
system. Systems without such mechanisms \cite{urgaonkar2002, 
bavier2004,wierman2003} typically assume that task values
(i.e., their importance) are the same, or are inversely proportional
to the resources required, or are set by an omniscient administrator.
However, in many cases, task values vary significantly, are not
correlated to resource requirements, and are difficult and
time-consuming for an administrator to set. Instead, market-based
resource allocation systems \cite{waldspurger1992, chun2000,
wellman2001,auyoung2004} rely on users to set the values
of their own jobs and provide a mechanism to encourage users to
truthfully reveal those values.

Despite these advantages, we are not aware of any currently
operational market-based resource allocation systems for computational
resources. We believe one key impediment is that previously proposed
systems impose a significant burden on users: frequent interactive
bidding, or, conversely, infrequent bidding that increases the latency
to acquire resources. Most users would prefer to run their program as
they would without a market-based system and forget about it until it
is done. The latency to acquire resources is important for
applications like a web server that needs to allocate resources
quickly in reaction to unexpected events (e.g., breaking news stories
from CNN). In addition, many market-based systems rely on a
centralized market that limits reliability and scalability.

In this paper, we present the \emph{Tycoon} distributed, market-based
resource allocation system. Each providing Tycoon host runs an
\emph{auctioneer} process that multiplexes the local physical
resources for one or more virtual hosts (using Linux VServers
\cite{vserver2004}). As a result, if an auctioneers fails, users can
still acquire resources at other hosts. Clients request resources from
auctioneers using \emph{continuous} bids that can be as infrequent as
the user wishes while still allowing immediate acquisition of
resources.

The contribution of this paper is the design, implementation, and
evaluation of Tycoon. We describe a prototype implementation of our
design running on a 22-host cluster distributed between Palo Alto in
California and Bristol in the United Kingdom. Tycoon can reallocate
all of the hosts in this cluster in less than 30 seconds. We show that
Tycoon encourages efficient usage of resources even when users make no
explicit bids at all. We show that Tycoon provides these benefits with
little overhead. Running a typical task on a Tycoon host incurs a less
than a 5\% overhead compared to an identical non-Tycoon host. Using
our current modest server infrastructure (450 MHz x86 CPU, 100 MB/s
Ethernet), limited tests indicate that our current design scales to
500 hosts and 24 simultaneous active users (or any other combination
with a product of 12,000). The main limitation of this implementation
is that it only manages CPU cycles (not memory, disk, etc.), but we
expect to resolve this by upgrading the virtualization software.

The paper is organized as follows. In \Section~\ref{sec:design_overview}, we
give an overview of the \emph{Tycoon} design. In
\Section~\ref{sec:architecture}, we describe the Tycoon architecture
in detail. In \Section~\ref{sec:experiments}, we present the results
of experiments using the Tycoon system. In
\Section~\ref{sec:relatedwork}, we review related work in resource
allocation. We describe some extensions to the basic design in
\Section~\ref{sec:FutureWork} and conclude in \Section~\ref{sec:conclusion}.

\section{Design Overview}
\label{sec:design_overview}

In this section, we present the service model and interface that
Tycoon provides to users. We describe the architecture of Tycoon in
more detail in \Section~\ref{sec:architecture}. 

\subsection{Service Model Abstraction}

The purpose of Tycoon is to allocate compute resources like CPU
cycles, memory, network bandwidth, etc. to users in an
\emph{economically efficient} way. In other words, the resources are
allocated to the users who value them the most. To give users an
incentive to truthfully reveal how much they value resources, users
use a limited budget of \emph{credits} to bid for resources. The form
of a bid is $(h, r, b, t)$, where $h$ is the host to bid on, $r$ is
the resource type, $b$ is the number of credits to bid, and $t$ is the
time interval over which to bid. This bid says, ``I'd like as much of
$r$ on $h$ as possible for $t$ seconds of usage, for which I'm willing
to pay $b$''.  This is a \emph{continuous} bid in that it is in effect
until cancelled or user runs out of money. 

The user submits this bid to the auctioneer that runs on host $h$.
This auctioneer calculates $b^r_i/t^r_i$ for each bid $i$ and resource
$r$ and allocates its resources in proportion to the bids. This is a
``best-effort'' allocation in that the allocation may change as other
bids change, applications start and stop, etc. Credits are not spent
at the time of the bid; the user must utilize the resource to burn the
credits. To do this, a user uses \texttt{ssh} to run a program. The
$t$ seconds of usage can be used immediately or later and at the same
time or in pieces, as the user wishes.

Note that the auctioneers are completely independent and do not share
information. As a result, if a user requires resources on two separate
hosts, it is his responsibility to send bids to those two markets.
Also, markets for two different resources on the same host are
separate.

This service model has two advantages. First, the continuous bid
allows user agents to express more sophisticated preferences because
they can place different bids in different markets. Specific
auctioneers can differentiate themselves in a wide variety of
ways. For example, an auctioneer could have more of a resource
(e.g. more CPU cycles), better quality-of-service (e.g., a guaranteed
minimum number of CPU cycles), a favorable network location, etc. A
user agent can compose bids however it sees fit to satisfy user
preferences. Second, since the auctioneers push responsibility for
expressing sophisticated bids onto user agents, the core
infrastructure can remain efficient, scalable, secure, and
reliable. The efficiency and scalability are a result of using only
local information to manage local resources and operating over very
simple bids. The security and reliability are a result of independence
between different auctioneers.

\subsection{Interface}

\begin{table*}
\begin{tabular}{|l|p{3.5in}|}
\hline 
Command&
Action\tabularnewline
\hline
\hline 
\texttt{tycoon create\_account host0 10 10 10}&
Create an account on host0 with a bid of 10 initial credits for CPU
cycles, memory, and disk.\tabularnewline
\hline 
\texttt{tycoon fund host0 cpu 90 1000}&
Fund the account on host0 using 90 credits to be spent over 1000
seconds for CPU cycles.\tabularnewline
\hline 
\texttt{tycoon set\_interval host0 cpu 2000}&
Change bid interval on the account to 2000 for CPU cycles.\tabularnewline
\hline 
\texttt{tycoon get\_status host0}&
Get status of account including the current balance, current interval,
etc. for each of the resources.\tabularnewline
\hline
\end{tabular}
\label{tab:Interface}
\caption{\small This table shows the main Tycoon user commands. }
\end{table*}

In this section, we describe how a user uses the system. The interface
requirments are important because we believe the bidding requirements
of previous economic systems were burdensome for users.

Table~\ref{tab:Interface} lists the main Tycoon user commands. These
are currently implemented as a Linux command-line tool, but they could
easily be implemented in a graphical user interface. The first action
a user takes is to create an account on a providing host. This
notifies auctioneers that a user intends to bid on that host and makes
an initial bid. The bid interval defaults to 10,000,000 seconds so
that the user is unlikely to run out of money. Account creation only
needs to be done rarely (in most cases once) per user and host. Users
usually perform account creation, like the operations that follow, on
many hosts, so the command-line tool allows the same operation to be
performed on multiple hosts in parallel.

At this point, the user can \texttt{ssh} into hosts and run his
application. Users are not required to change their bids when they
start and stop tasks. They can do so to optimize their resource usage,
if they wish. However, the auctioneers will still deduct credits when
he runs. As a result, users who run infrequently will get more
resources than those who run continuously.  If the user chooses, he
can transfer more money to his account and/or change the bidding
interval. He might have a critical task for which he is willing to
spend credits at a higher rate, or, conversely, he might have a very
low priority job, for which he wishes to decrease his spending
rate. The key point is that the users are relieved from any mandatory
interaction with the system.

\section{Architecture}
\label{sec:architecture}

\emph{Tycoon} is split into the following components: service location
service (SLS), bank, auctioneer, and agent. The design of the SLS and
bank are not novel, but we describe them here because they are
necessary components for a working implementation.

\begin{figure*}[htb]
\begin{center}
\includegraphics[width=6in]{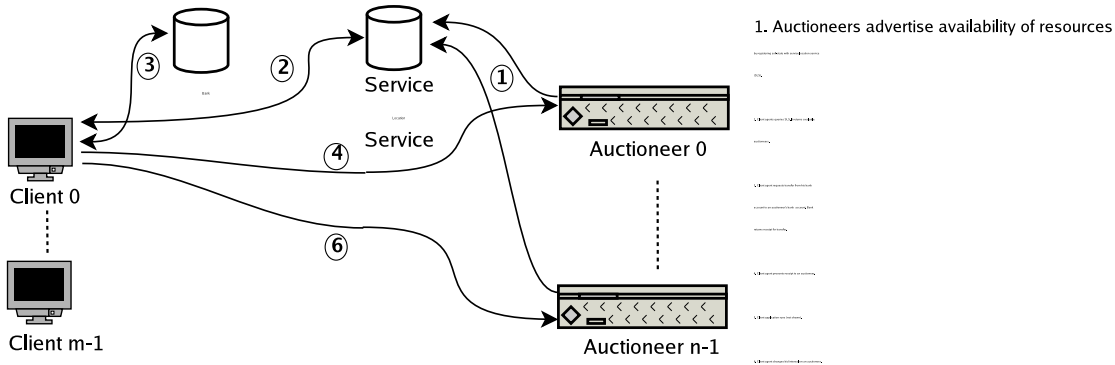}
\caption{\small This figure gives an overview of how the Tycoon
  components interact. }
\label{fig:protocoloverview}
\end{center}
\end{figure*}

\subsection{Service Location Service}

Auctioneers use the service location service to advertise resources,
and agents use it to locate resources (as shown in steps 1 and 2 in
Figure~\ref{fig:protocoloverview}). Our prototype uses a simple
centralized soft-state server, but the other components would work
just as well with more sophisticated and scalable service location
systems (e.g., Ganglia \cite{massie2004} and SWORD
\cite{oppenheimer2004}). Auctioneers register their status with the
SLS every 30 seconds and the SLS de-registers any auctioneer that has
not contacted it within the past 120 seconds. This status consists of
the total amount bid on the host for each resource, the total amount
of each resource type available (e.g., CPU speed, memory size, disk
space), etc. The status is cryptographically signed by the auctioneer
and includes the auctioneer's public key. Clients store this key and
use it to authenticate the results of later queries and also to
authenticate direct communications with the auctioneer.

The soft-state design allows the system to be robust against many
forms of hardware and software failures. The querying agents may
receive stale information from the SLS, but they will receive updated
information if they elect to contact an auctioneer directly.

\subsection{Bank}
\label{sec:bank}

The bank maintains account balances for all users and providers. Its
main task is to transfer funds from a client's account to a
provider's account (shown in step 3 in
Figure~\ref{fig:protocoloverview}). 

We assume that the bank has a well-known public key and that the bank
has the public keys of all the users. These are the same requirements
for any user to securely use a host with or without a
market-based resource allocation system. We further assume roughly
synchronized clocks. In describing the transfer protocol, we use Alice
and Bob as the fictional example sender and receiver. Alice begins by
sending a message to the bank as follows:
\[
Alice, Bob, amount, time, 
\]
\[
Sign_{Alice}(Alice, Bob, amount, time)
\]
$Sign_{Alice}$ is the DSA signature function using Alice's private
key. The bank verifies that the signature is correct, which implies
that the message is from Alice, that the funds are for Bob, and that
the amount and time are as specified. The bank keeps a list of recent
messages and verifies that this message is new, thus guarding against
replay attacks. Assuming this is all correct and the funds are
available, the bank transfers $amount$ from Alice to Bob and responds
with the following message (the \emph{receipt}):
\[
Alice, Bob, amount, time,
\]
\[
Sign_{Bank}(Alice, Bob, amount, time)
\]
The bank sends the same time as in the first message. Alice verifies
that the amount, time, and recipient are the same as the original
message and that the signature is correct. Assuming the verification
is successful, Alice forwards this message to Bob as described in
\Section~\ref{sec:Auctioneer}. Bob keeps a list of recent receipts and
verifies that this receipt is new, thus guarding against replay
attacks. 

The advantages of this scheme are simplicity, efficiency, and
prevention of counterfeiting. Micro-currency systems are generally
complex, have high overhead, and only discourage counterfeiting.  The
disadvantages of this approach are scalability and vulnerability to
compromise of the bank. However, bank operations are relatively
infrequent (see \Section~\ref{sec:SettingBids} for how bids can be
changed without involving the bank), so scalability is not a critical
issue for moderate numbers of users and hosts, as we show in
\Section~\ref{sec:NetworkOverhead}. The vulnerability to compromise of the
bank could be a problem and we discuss possible solutions in
\Section~\ref{sec:FutureWork}.

\subsection{Auctioneer}
\label{sec:Auctioneer}
Auctioneers serve four main purposes: management of local resources,
collection of bids from users, allocation of resources to users
according to their bids, and advertisment of the availability of local
resources.

\subsubsection{Virtualization}
To manage resources, an auctioneer relies on a virtualization system
and a local allocation system. Our implementation uses Linux VServer
(with modifications from PlanetLab) for virtualization. VServer
provides each user with a separate file system and gives the
appearance that he is the sole user of a machine, even if the physical
hardware is being shared. The user accesses this virtual machine by
using \texttt{ssh}.

VServers virtualize at the system call level, which provides the
advantage of low overhead.  We show in \Section~\ref{sec:HostOverhead}
that the total auctioneer overhead, including VServers, is at most ten
percent and usually much less. Systems that virtualize at the
hardware level like VMWare \cite{vmware2004} or Disco
\cite{bugnion1997} have significantly more overhead
\cite{dragovic2003}.

For local allocation, Tycoon uses the plkmod proportional share
scheduler \cite{bavier2004}, which implements the standard
proportional share scheduling abstraction \cite{tijdeman1980}. The
disadvantage of VServers and plkmod is that they do not completely
virtualize system resources. This is why Tycoon currently 
only manages CPU cycles. In \Section~\ref{sec:FutureWork} we discuss
new virtualization and allocation systems that provide this
functionality.

\subsubsection{Setting Bids}
\label{sec:SettingBids}
The second purpose of auctioneers is to collect bids from users.
Auctioneers store bids as two parts for each user: the local 
account balance, and the bidding interval. The local balance is the amount of
money the user has remaining locally. The bidding interval specifies
the number of seconds over which to spend the local balance. Users
have two methods of changing this information: {\tt fund} and
{\tt set\_interval}.  {\tt fund} transfers money from the user's bank
account to the auctioneer's bank account, and conveys that fact to the
auctioneer. It has the disadvantage that it requires significant
latency (100 ms) and it requires communication with the bank, which
may be offline or overloaded.  {\tt set\_interval} sets the bidding
interval at the auctioneer without changing the local balance. It only
requires direct communication between the client and the auctioneer,
so it provides a low latency method of adjusting the bid until the
local balance is exhausted.

In describing the {\tt fund} protocol, we again use Alice and Bob as
examples. We assume that Alice and Bob already have each other's
public keys and that Alice has the value $nonce_{Alice}$.
A nonce is a unique token which Bob has never seen from Alice before. 
In the current implementation it is an increasing counter.
First, Alice gets a bank receipt as
described above. She then sends the following message to Bob:
\[
Alice, Bob, nonce_{Alice}, interval, receipt,
\]
\[
Sign_{Alice}(Alice, Bob, nonce_{Alice}, interval, receipt)
\]
The nonce allows Bob to detect replay attacks. Bob verifies that he is
the recipient of this message, that the nonce has not been used
before, that the receipt specifies that Alice has transferred money
into his account, that the bank has correctly signed the receipt, and
that Alice has correctly signed this message. Assuming this is all
correct, Bob increases Alice's local balance by the amount specified
in the receipt and sets Alice's bidding interval to
$interval$.  {\tt set\_interval} is identical, except that it does not
include the bank receipt.

The key advantage of separating {\tt fund} and {\tt set\_interval} is that
it reduces the frequency of bank operations. Users only have to fund
their hosts when they wish to change the set of hosts they are running
on or when they receive income. For most users and applications, we
believe this is on the order of days, not seconds. Between fundings,
users can modify their bids by changing the bidding interval, as
described in the next section.

\subsubsection{Allocating Resources}
\label{sec:ResourceAllocation}

The third and most important purpose of auctioneers is to use
virtualization and the users' bids to allocate resources among the
users and account for usage. Although our current implementation only
allocates CPU cycles because of virtualization limitations, the
following applies to both rate-based (e.g., CPU cycles and network
bandwidth) and space-based (e.g., physical memory and disk space)
resources. In addition, we initially describe a proportional
share-based function, but there are other allocation functions with
desirable properties (e.g., Generalized Vickrey Auctions, described below).

For each user $i$, the auctioneer knows the local balance $b_i$ and
the bidding interval $t_i$. The auctioneer calculates the bid as
$b_i/t_i$. Consider a resource with total size $R$ (e.g., the number of
cycles per second of the CPU or the total disk space) over some period
$P$.  The allocation function for $r_i$, the amount of resource
allocated to user $i$ over $P$, is
\[
\label{eq:allocationfunc}
r_i = \frac{\frac{b_i}{t_i}}{{\sum_{j=0}^{n-1} \frac{b_j}{t_j}}}R.
\]
Let $q_i$ be the amount of the resource that $i$ actually consumes
during $P$, then the amount that $i$ pays per second is 
\[
s_i = min\left(\frac{q_{i}}{r_{i}},1\right)  \frac{b_{i}}{t_{i}}. 
\]
This allows users who do not use their full allocation to pay less
than their bid, but in no case will a user pay more than his bid.

There are a variety of implementation details. First, the auctioneer
gets the number of cycles used by each user from the kernel to
determine if $q_i < r_i$. Second, we set $P=10s$, so the auctioneer
charges users and recomputes their bids every 10 seconds. This value
is a compromise between the overhead of running the auctioneer and the
latency in changing the auctioneer's allocation. With tighter
integration with the kernel and the virtualization system, $P$ could
be as small as the scheduling interval ($10ms$ on most
systems). Third, users whose bids are too small relative to the other
users are logged off the system. Users who bid for less than .1\% of
the resource would run infrequently while still consuming overhead for
context-switching, accounting, etc., so the auctioneer logs them off, starting
with the smallest bidder.

The advantages of this allocation function (\ref{eq:allocationfunc})
are that it is simple, it can be computed in $O(n)$ time, where $n$ is
the number of bidders, it is fair, and it can be optimized across
multiple auctioneers by an agent (described in
\Section~\ref{sec:Agent}). It is fair in the sense that all users who
use their entire allocation pay the same per unit of the resource.

The disadvantage is that it is not \emph{strategyproof}. In the simple
case of one user running on a host, that user's best (or
\emph{dominant}) strategy is to make the smallest possible bid, which
would still provide the entire host's resources. If there are multiple
users, then the user's dominant strategy is to bid his
valuation. Since, the user's dominant strategy depends on the actions
of others, this mechanism is not strategyproof. One possible
strategyproof mechanism is a Generalized Vickrey Auction (GVA)
\cite{varian1995}. However, this requires $O(n^2)$ time, it is not
fair in the sense described above, and it is not clear how to optimize
bidding across multiple GVA auctioneers.

\subsubsection{Advertising Availability}

The auctioneer must advertise the availability of local resources so
that user agents can decide whether to place bids. For each resource
available on the local host, the auctioneer advertises the total
amount available, and the total amount spent at the last
allocation. In other words, the auctioneer reports 
\[
\displaystyle\sum_{j=0}^{n-1} s_i.
\]
This may be less than the sum of the bids because some tasks did not
use their entire allocation. We report this instead of the sum of the
bids because it allows the agent to more accurately predict the cost
of resources (as required the algorithm described in
\Section~\ref{sec:BestResponseAlgorithm}). Note that this information
allows agents to make appropriate bids without revealing the exact
amounts of other users' individual bids. Revealing that information
would allow users to know each other's valuations, which would allow
gaming the auctions.

\subsection{Agent} 
\label{sec:Agent}

The role of a tycoon agent is to interpret a user's preferences,
examine the state of the system, make bids appropriately, and verify
that the resources were provided. The agent is involved in steps 2, 3,
4, and 6 of Figure~\ref{fig:protocoloverview}. Given the diversity of
possible preferences, we chose to separate agents from the
infrastructure to allow agents to evolve independently. This is a
similar approach to the end-to-end principle used in the design of the
Internet \cite{cerf1974,clark1988,saltzer1984}, where
application-specific functionality is contained in the end-points
instead of in the infrastructure. This allows the infrastructure to be
efficient, while supporting a wide variety of applications.


There are a wide variety of preferences that a user can specify to his
agent. Tycoon provides for both high-level preferences that an agent
interprets and low-level preferences that users must specify in
detail. Examples of high level preferences are wanting to maximize the
expected number of CPU cycles or to seek machines with a minimum
amount of memory, or some combination of those preferences. Tycoon
allows uncertainty in the exact amount of resource received because
other applications on the same host may not use their allocation
and/or other users may change their bids.

\subsubsection{Best Response Algorithm}
\label{sec:BestResponseAlgorithm}
In a system with many machines, it is very difficult for users to bid
on individual machines to maximize their utilization of the system.
In Tycoon, we allow the user to only specify the total bids, or the
budget, he is willing to spend and let the agent compute the bids
on the machines to maximize the user's utility.  In order to compute
the optimum bids, the agent must first know the user's utility as
a function of the fraction of the machines assigned to the user. Since
it is difficult, if not impossible, to figure out the exact
formulation of the utility function, we assume a linear utility
function for each user.  That is, each user specifies a non-negative
weight for each machine to express his preference of the machine.
Such a weight is chosen by the user and determined mainly by two
factors: the system configuration and the user's need.  They may vary
from user to user.  For example, one user may have higher weight on
machine $A$ because it has more memory, and another user may have
higher weight on $B$ because it has a faster CPU.  The weights are kept
private to the users.

Now, suppose that there are $n$ machines, and a user has weight $w_i$
on machine $i$ for $1\leq i\leq n$.  If the user gets fraction $r_i$
from machine $i$, then his utility is
\[U =\sum_{i=1}^{n} w_i r_i\,.\]

The agent's goal is to maximize the user's utility under a given
budget, say $X$, and the others' aggregated bids on the machines.
Suppose that $y_i$ is the total bid by other users on machine $i$.  The user's
share on $i$ is then $\frac{x_i}{x_i+y_i}$ if he bids $x_i$ on machine $i$.
Therefore, the agent needs to solve the following optimization
problem:
\[\mbox{maximize } \sum_{i=1}^n w_i \frac{x_i}{x_i+y_i}\,,\quad\mbox{s.t.}\]\label{eqn:x}
\[x_i\geq 0\,,\quad\mbox{for $1\leq i\leq n$, and}\]
\[\hspace*{-2.3cm}\sum_{i=1}^n x_i = X\,.\]

This optimization problem can be solved by using the following algorithm. 
\begin{enumerate}
\item sort $\frac{w_i}{y_i}$ in decreasing order, and suppose that 
\[\frac{w_1}{y_1}\geq\frac{w_2}{y_2}\geq\cdots\geq\frac{w_n}{y_n}\,.\]

\item compute the largest $k$ such that
\[\frac{\sqrt{w_ky_k}}{\sum_{j=1}^k \sqrt{w_jy_j}}(X+\sum_{j=1}^k y_j)-y_k\geq 0\,.\]

\item set $x_i=0$ for $i>k$, and for $1\leq i\leq k$, set
\[x_i=\frac{\sqrt{w_iy_i}}{\sum_{j=1}^k \sqrt{w_jy_j}}(X+\sum_{j=1}^k y_j)-y_i\,.\]
\end{enumerate}

The above algorithm takes $O(n\log n)$ time as sorting is the most
expensive step.  It is derived by using Lagrangian multiplier method.
Intuitively, the optimum is achieved by the bids where the bid on each
machine has the same marginal value.  The challenge is to select the
machines to bid on.  Roughly speaking, one should prefer to bid on a
machine if it has high weight on the machine and if other's bids on
that machine is low.  That is the intuition behind the first sorting
step. We omit the correctness proof of the algorithm due to the space
limitation.

One problem with the above algorithm is that it spends the entire
budget.  In the situation when there are already heavy bids on the
machines, it might be wise to save the money for later use.  To deal
with the problem, a variation is to also prescribe a threshold
$\lambda$ to the agent and require that the margin on each machine is
not lower than $\lambda$, in addition to the budget constraint. Such
problem can be solved by an easy adaptation of the algorithm.

\subsubsection{Predictability}

Instead of maximizing its expected value, some applications may prefer
to maintain a minimum amount of a resource. An example of this is
memory, where an application will swap pages to disk if it has less
physical memory than some minimum, but few applications benefit
significantly from having more than that. Tycoon allows agents to
express this preference by putting larger bids on fewer machines. Let
$R$ be the total resource size on a host and $B$ be the sum of the
users' bids for the resource, excluding user $i$. From
(\ref{eq:allocationfunc}), the user $i$'s agent can compute that to
get $r_i$ of a resource, it should bid
\[
b_i = \frac{r_iB}{R-r_i}.
\]
However, this only provides an expected amount of $r_i$. To provide
higher assurances of having this amount, the agent bids more than
$b_i$. To determine how much more, the agent maintains a history of
the bids at that host to determine the likelihood that a particular
bid will result in obtaining the required amount of a resource.
Assuming that the application only uses $r_i$ of the resource, the
user will pay more per unit of the resource than if his agent had just
bid $b_i$ (see \Section~\ref{sec:ResourceAllocation}), but that is the
price of having more predictability.

\subsubsection{Scalability}
Since the computational overhead of the agent is low, the main
scalability concern is communications overhead. When making bids, a
user agent may have to contact a large number of auctioneers, possibly
resulting in a large queueing delay. For example, to use 100 hosts,
the agent must send 100 messages. Although the delay to do this is
proportional to the amount of resources the user is using, for very
large numbers of hosts and a slow and/or poorly connected agent host,
the delay may be excessive. In this case, the agent can use an
application-layer multicast service (e.g., Bullet \cite{kostic2003})
to reduce the delay. Since changing a bid consists of simply setting
an interval, the user agent can use a multicast service to send out
the same interval to multiple auctioneers. This would essentially make
the communication delays logarithmic with respect to number of hosts.

\subsubsection{Verification}

One potential problem with all auction-based systems is that
auctioneers may cheat by charging more for resources than the rules of
the auction dictate. However, one advantage of Tycoon is that it is
market-based so users will eventually find more cost-effective
auctioneers. Cost-effectiveness is an application-specific metric.
For example, an application may prefer a slow host because it has a
favorable network location. Users who are interested in CPU cycles
would view that as a host with poor cost-effectiveness.  However, in
many applications, the agent can measure cost-effectiveness fairly
accurately. As an example, the rendering application we use in
\Section~\ref{sec:experiments} uses frames rendered per second as its
utility metric. As a result, the cost-effectiveness is frames rendered
per second per credit spent for each host.

The measured cost-effectiveness is then used as the host weight for
the best-response algorithm. This algorithm will automatically drop a
host from bidding when it sees that it is significantly less
cost-effective than the others. Effectively, Tycoon treats a cheating
host as a host with poor cost-effectiveness. Therefore we do need
sophisticated techniques to detect or prevent cheating. If no agents
wants to spend credits at a cheating auctioneer, the monetary
incentive to cheat is greatly reduced.

\subsection{Funding Policy}
Funding policy determines how users obtain funds. We define \emph{open
loop} and \emph{closed loop} funding policies. In an open loop funding
policy, users are funded at some regular rate. The system
administrators set their income rate based on exogenously determined
priorities. Providers accumulate funds and return them to the system
administrators. In a closed loop (or \emph{peer-to-peer}) funding
policy, users themselves bring resources to the system when they
join. They receive an initial allotment of funds, but they do not
receive funding grants after joining. Instead, they must earn funds by
enticing other users to pay for their resources. A closed loop funding
policy is preferable because it encourages service providers to
provide desirable resources and therefore should result in higher
economic efficiency.

\section{Experiments}
\label{sec:experiments}

\subsection{Experimental Setup}
\label{sec:ExperimentalSetup}

\begin{table*}
\small
\begin{center}
\begin{tabular}{|c|c|c|c|c|c|}
\hline
Processor Variety & CPU & Memory & Disk & \# nodes & Location \\
\hline
\hline
Pentium III & 1 GHz & 2 GB & 32 GB SCSI & 4 & US \\
\hline
Mobile Pentium III & 900 MHz & 512 MB & 40 GB IDE & 8 & UK \\
\hline
Pentium III & 550 MHz & 256 MB & 10 GB IDE & 2 & US \\
\hline
Pentium II & 450 MHz & 128 MB & 10 GB IDE & 6 & UK \\
\hline
\end{tabular}
\end{center}
\label{tab:Cluster}
\caption{Specifications of the four types of computers used in the test cluster.
}
\end{table*}

The experiments in this section were run the on the hosts shown in
Table~\ref{tab:Cluster}. A were running Linux with the PlanetLab
2.4.22 kernel, which includes VServer and plkmod. 

\subsection{Agility}
\label{sec:Agility}

In this section, we report the results of experiments to test agility,
the ability to adapt to changes in demand. As a workload, we used the
Maya~6.0 image rendering software to renderer frames in a movie scene.
The jobs were dispatched using the Muster job queue, an off-the-shelf
product that manages distributed rendering jobs. During the experiment
two users were rendering concurrently on each node.

\begin{figure}[htb]
\includegraphics[angle=0, width=\columnwidth]{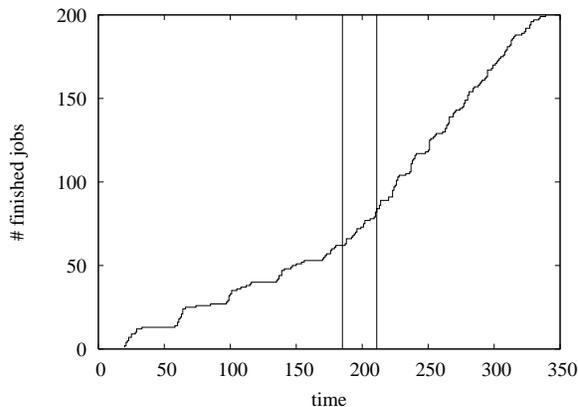}
\caption{\small This figure shows a user increasing his share at 190
seconds by decreasing the bidding interval.  As a result, the
throughput increases by 210 seconds. }
\label{fig:changebid}
\end{figure}

First, we examine the time for a user to acquire more resources to
finish his rendering job sooner. In Figure~\ref{fig:changebid}, a user
has initialized his nodes with \$10 to be spent over 30,000 seconds.
He submits a 200 frame rendering job to the Tycoon cluster.  Someone
else is already running on the cluster. Using the bids of both users,
auctioneers allocate the new user about twenty percent of each node.
After running for three minutes, the user notices that the job is not
likely to finish early enough, so he changes the spending interval to
300 seconds on all nodes.  This will leave him with fewer credits at
the end of the run than if he left the interval at 30,000, but it is
worth it to him. The time at when he changes the interval is marked by
the left vertical line. About twenty seconds later, the right vertical
line marks the time at which the user is able to detect an increased
rate of rendering.  Afterward, the frames finish at an increased
speed, and the job finishes on time.

This demonstrates the system's ability to quickly reallocate
resources. As in this case, this could be because a user cannot
accurately estimate the resource requirements of his application.
Other possible causes are that hosts have failed, the load has
increased, the user's deadline has changed, etc. The agility of the
system allows users to compensate for uncertainty. 

\begin{figure}[htb]
\includegraphics[angle=0, width=\columnwidth]{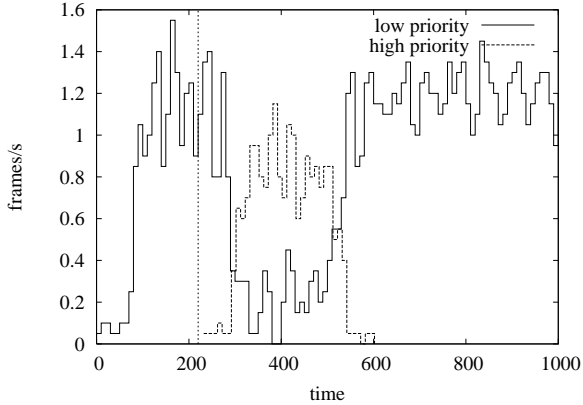}
\caption{\small This figure shows a low priority job with a small
share getting lower throughput when a high priority job arrives. }
\label{fig:rateplot}
\end{figure}

In a second experiment, we examine the system's ability to change
allocations when a high priority job is started.  In this scenario,
two users are rendering on the cluster.  One user performs a low
priority render, and he funds his nodes with \$10 for 100,000 seconds.
A second user funds his nodes with \$10 for 10,000 seconds.
Initially, only the low priority job is running, but after 220
seconds, the second user submits a rendering job to the system.

Figure~\ref{fig:rateplot} shows the average rate at which frames are
finished for the two jobs. First the low priority job runs alone, at
an average rate of 1.1 frames per second.  When second users submits
the high priority job, (marked with a vertical line in the figure), the
throughput of the low priority job decreases almost immediately to 0.2
frames per second, and the high priority job starts to render at 0.9
frames per second.  As soon as the high priority job finishes, the low
priority job starts to utilize the CPUs again, and gets an increased
throughput. 

When high priority job first starts, it has lower throughput because
it is waiting for disk I/O.  During that time, the low priority job is
able to continue to utilize the CPU.  As soon as the high priority job
is ready to run, it produces frames at almost full speed.  Based on
the bids, its share is 90\%.  The actual throughput is on average
0.9/1.1, which is slightly lower.  This is also because of disk I/O
delays.  The throughput penalty from I/O is higher for the high
priority task than for the low priority task because it issues more
I/O operations. This is an artifact of our version of VServer being
unable to regulate disk I/O bandwidth. If our virtualization layer had
that capability, the actual throughput would be closer to the ideal of
90\%.

\begin{figure}[htb]
\includegraphics[angle=0, width=\columnwidth]{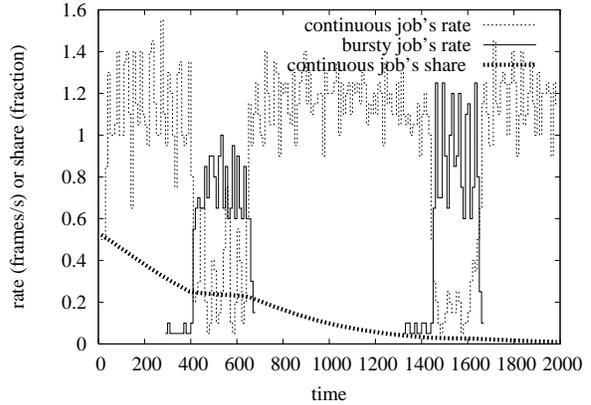}
\caption{\small This figure shows how a user that runs infrequently
can receive more resources when he does run in comparison to a user
that runs continuously. }
\label{fig:bursty}
\end{figure}

In the third experiment, we show how the system treats a user who runs
infrequently in comparison to one that runs continously. Both users
initialize their nodes with \$10 for 300 seconds.  One user starts a
long continuous job on the cluster.  While the user is running alone,
his share decreases in proportion to $(1-P/t_i)^{\tau/P}$ where
$\tau$ is the time since the start of the experiment, $t_i=300$ is the
funding interval, and $P=10$ is the auctioneers' update interval.
Since the infrequent user is not running, the continuous job initially
gets to use the whole cluster, as shown in Figure~\ref{fig:bursty}.






After 400 seconds, the infrequent user starts running.  Since it has
not spent any money, it's share is 75 percent, and the job that has
been running has a 25 percent share. Since both jobs continue to pay
in proportion to their balance, their shares remain at 75 and 25
percent, respectively, until the infrequnt users stops running. The
infrequent user returns at 1300 seconds and again he gets most of the
resources. In this case, he gets most of the resources because the
continuous user's share has dropped considerably. 

The key point about this result is that the system encourages
efficient usage of resources even when users do not make explicit
bids. In this experiment, the users bids were identical, which could
have been set when their accounts were created. Despite this, the
infrequent user is rewarded for being judicious in his resource
consumption, while the continuous user is penalized for running all
the time. In comparison, a proportional share system would allocate
50\% of the resources to each user when both are running. This gives
no disincentive for the continuous user to stop running. The
performance improvement for the infrequent user is
$(.75-.50)/.50=50\%$ for one continuous user. For $n$ continuous users,
the performance improvement is $(.75-(1/n))/(1/n)$, which goes to
infinity as $n$ goes to infinity. 

\subsection{Host Overhead}
\label{sec:HostOverhead}

\begin{figure}
\includegraphics[width=\columnwidth]{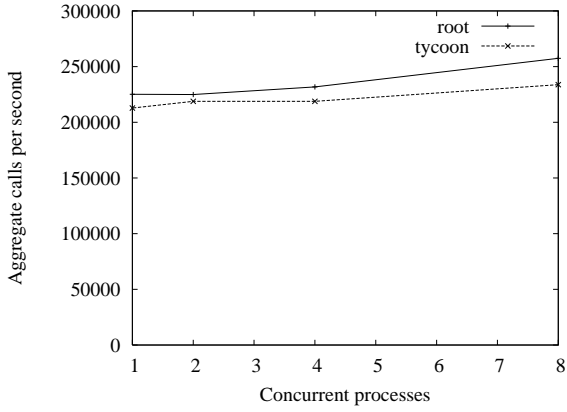}
\caption{System Call Performance}
\label{fig:syscall}
\end{figure}

\begin{figure}
\includegraphics[width=\columnwidth]{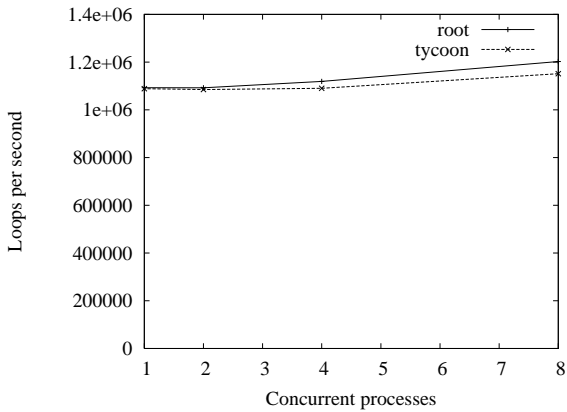}
\caption{CPU-bound Task Performance}
\label{fig:dhry}
\end{figure}

\begin{figure}
\includegraphics[width=\columnwidth]{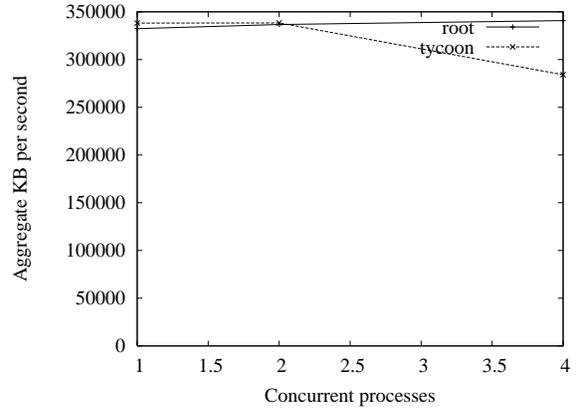}
\caption{Disk Read Performance}
\label{fig:read}
\end{figure}

\begin{figure}
\includegraphics[width=\columnwidth]{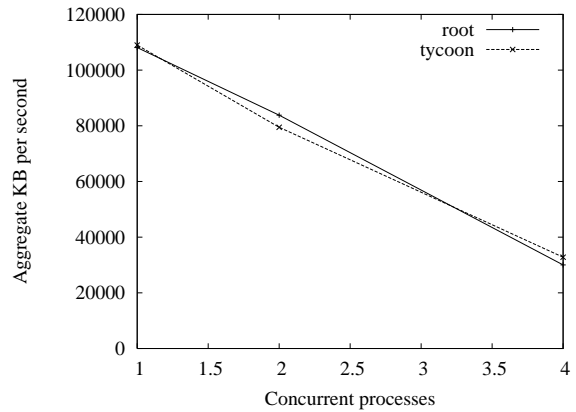}
\caption{Disk Write Performance}
\label{fig:write}
\end{figure}

This set of experiments measures the overhead incurred by using Tycoon
rather than using the same Linux computer without Tycoon. This
overhead includes VServer, plkmod, and the auctioneer overhead. We
compared this relative performance for three distinct types of
operations. They are illustrated in Figures~\ref{fig:syscall},
\ref{fig:dhry}, \ref{fig:read} and \ref{fig:write} for system call
overhead, CPU-bound computation, disk reading and disk writing,
respectively. In these experiments, from one to eight programs
designed to test a particular type of operation are invoked
simultaneously by \texttt{ssh}. For the Tycoon experiments, each
program is started as a distinct user. In the root scenario, the
programs are all run as root. The sum of the scores of all of the
concurrent processes is plotted as a function of the number of
concurrent processes. 

For CPU-bound processes and for a few I/O-bound processes, Tycoon has
less than five percent overhead. We expect the bulk of cluster
applications to be similar to these micro-benchmarks. For processes
that involve many system calls, the overhead is capped at ten percent,
but we do not expect many Tycoon processes to be system
call-heavy. The overhead for Tycoon is most significant for many disk
reading processes. This may be due to the additional memory overhead
of VServer reducing the size of the buffer cache, but we are still
investigating this.



\subsection{Network Overhead}
\label{sec:NetworkOverhead}

\begin{table}
\small
\begin{tabular}{|l|l|l|l|l|}
\hline
Operation & SLS & Bank & Auc. & Agent \\
\hline
\hline
Registration (per min.) & 260 & & 9634 &  \\
\hline
SLS query (20 hosts) & 89K & & & 311 \\
\hline
Bank transfer & & 1198 & & 610 \\
\hline
Account creation & & & 5901 & 3592 \\
\hline
Spending rate change & & & 793 & 719 \\
\hline
\end{tabular}
\caption{Bytes sent from the specified entity while conducting the specified operation.}
\label{tab:NetworkOverhead}
\end{table}

The primary bottlenecks that prevent the unfettered scaling of a
Tycoon cluster are the two centralized servers, the service location
server (SLS) and the bank. Table~\ref{tab:NetworkOverhead} quantifies
the costs of performing the most common operations on a Tycoon
cluster.

The most frequent process is the maintenance of soft-state between the
auctioneers and the SLS. Assuming that the SLS is allowed to use
100Mb/s network bandwidth (e.g., it is on a 1Gb/s network), it can
manage up to 75,000 Tycoon hosts. If clients use the best response
agent to operate on the Tycoon cluster, they must issue repeated
host-list queries to the SLS to compute their optimal bidding
strategy. If the agent updates its strategy once a minute, it costs
roughly 4KB/minute per agent per host. Again assuming this task is
allocated 100Mb/s of bandwidth, the product of the number of agents
and number of hosts must not exceed 187M. Hence assuming that there
are 75K hosts in the cluster, there may be up to 2500 agents running
concurrently. Similarly if there are only 2500 hosts, there may be up
to 75K agents.

A less frequent operation is bank transfers from users to hosts. This
task depends less on bandwidth and more on the speed of the bank
system in performing large integer arithmetic for authentication. On a
450 MHz Pentium III, this operation requires an average of
100ms. Assuming user perform bank operations every twenty minutes per
user per host, this bank supports an active user-host product of
12,000, which would allow 24 simultaneous active users on a 500 host
cluster. As a result, for the immediate future, we do not believe a
centralized bank is a significant problem. One reason is that much
faster hardware is available. A 3 GHz bank should support 6.7 times
the number of users or hosts or combination thereof. Another reason is
that the current protocol performs only one credit transfer per
connection. It could be optimized to perform multiple transfers per
connection which would amortize the authentication and communication
costs. Finally, twenty minutes is a very conservative estimate of bank
operations. A more likely frequency is once a day. This would allow
even the current slow hardware and unoptimized protocol to support a
user-host product of 864000. A centralized bank is not likely to limit
scalability in practice.

\section{Related Work}
\label{sec:relatedwork}

In this section, we describe related work in resource allocation.
There are two main groups: those that incorporate an economic
mechanism\footnote{By \emph{mechanism} we mean the system that
provides an incentive for users to reveal the truth (e.g., an
auction)}, and those that do not.

One of the key non-economic abstractions for resource allocation is a
computer science context is Proportional Share (PS), originally
documented by Tijdeman \cite{tijdeman1980}. Each PS process $i$ has a
weight $w_i$. The share of a resource that process $i$ receives over
some interval $t$ where $n$ processes are running is
\begin{equation}
\label{eq:proportionalshare}
\frac{w_i}{{\displaystyle\sum_{j=0}^{n-1} w_j}}.
\end{equation}
PS maximizes utilization because it always provides resources to needy
processes. One problem is that PS is usually applied by giving each
user a weight and directly transferring that weight to the user's
processes. However, a user may not weigh all of his processes equally
and PS does not give an incentive for users to differentiate his
processes. As a result, as a system becomes more loaded, the low value
processes consume more resources, until the high value processes
cannot make useful progress (as shown by Lai, et al. \cite{lai2004}). 

One common method for dealing with this problem is to rely on social
mechanisms to set the PS weights appropriately. A system administrator
could set them based on input from users or users could ``horse
trade'' high weights amongst themselves. Although these mechanisms
work well for small groups of people that trust each other, they do
not scale to larger groups and they have a high overhead in user
time. 

Most recent work by Waldspurger and Weihl \cite{waldspurger1994},
Stoica, et al. \cite{stoica1996}, and Nieh, et al. \cite{nieh2001} on
PS has focused on computationally efficient and fair
implementations. Lottery scheduling \cite{waldspurger1994} is a 
PS-based abstraction that is similar to the economic approach in that
processes are issued tickets that represent their
allocations. Sullivan and Seltzer \cite{sullivan2000} extend this to
allow processes to barter these tickets. Although this work provides
the software infrastructure for an economic mechanism it does not
provide the mechanism itself.

Similarly, SHARP (described by Fu, et al. \cite{fu2003}) provides the
distributed infrastructure to manage tickets, but not the mechanism or
agent strategies. In addition, SHARP and work by Urgaonkar, et
al. \cite{urgaonkar2002} use an overbooking resource abstraction
instead of PS. An overbooking system promises probabilistic resources
to applications. Tycoon uses a similar abstraction for applications
that require a minimum amount of a resource.

Another class of non-economic algorithms examine resource allocation
from a scheduling (surveyed by Pindedo \cite{pinedo2002}) perspective
using combinatorial optimization (described by Papadimitriou and
Steiglitz \cite{papadimitriou1982}) or by examining the resource
consumption of tasks (a recent example is work by Wierman and
Harchol-Balter \cite{wierman2003}). However, these assume that the
values and resource consumption of tasks are reported accurately. This
assumption does not apply in the presense of strategic users. We view
scheduling and resource allocation as two separate functions. Resource
allocation divides a resource among different users while scheduling
takes a given allocation and orders a user's jobs.

Examples the economic approach are Spawn (by Waldspurger, et
al. \cite{waldspurger1992}), work by Stoica, et
al. \cite{stoica1995}., the Millennium resource allocator (by Chun, et
al. \cite{chun2000}), work by Wellman, et al. \cite{wellman2001}, and
Bellagio (by AuYoung, et al. \cite{auyoung2004}).

Spawn and the work by Wellman, et al. uses a reservation abstraction
similar to the way airline seats are allocated. Although reservations
allow low risk, the utilization is also low because some tasks do not
use their entire reservations.  Service applications (e.g., web
serving, database serving, and overlay network routing) result in
particularly low utilization because they typically have bursty and
unpredictable loads. Another problem with reservations is that they
can significantly increase the latency to acquire resources.  A
reservation by one user prevents another user from using the resources
for the duration of the reservation, even if the new user is willing
to pay much more for the resources than the first user. Reservations
are typically on the order of minutes or hours (Spawn used 15
minutes), which is too much delay for a highly bursty and
unpredictable application like web serving.

The proportional share abstraction used in the Millennium resource
allocator comes the closest to that used in Tycoon. We extend that
abstraction with continuous bids, the best-response agent algorithm,
and secure protocols for bidding.

Bellagio uses a centralized allocator called SHARE developed by Chun,
et al. \cite{chun2004}. SHARE takes the combinatorial auction approach
to resource allocation. This allows users to express preferences with
complementarities like wanting host A and host B, but not wanting host
A without B or B without A. The combinatorial auction approach relies
on a centralized auctioneer to guarantee that the user either gets
both A and B or else nothing. Economic theory predicts that solving
this NP-complete problem provides an allocation with optimal economic
efficiency. Tycoon addresses the combinatorial problem in a possibly
less economically efficient, but more scalable way. In Tycoon, credits
are only spent when the user actually consumes resources, so the
user's agent can see that it only has A before his application runs
and thereby prevent wasting credits on an unvalued resource.  The
disadvantages of the combinatorial auction approach are the
centralized auctioneer and the difficulty of the combinatorial auction
problem. The centralized auctioneer is vulnerable to compromise and
limits the scalability of the system, especially since it must be
involved in all allocations. Moreover, even computationally efficient
heuristic algorithms operate on the order of minutes, while Tycoon
reallocates in less than ten seconds. Recent work by Hajiaghayi
\cite{hajiaghayi2004} on online resource allocation may be able to
reduce the delay of the combinatorial approach.

\section{Future Work}
\label{sec:FutureWork}

One area of future work is more complete virtualization. Our prototype
implementation uses early versions of VServer and plkmod which only
support virtualization of CPU cycles. Later versions of VServer, Xen
\cite{dragovic2003}, and the Class-based Kernel Resource Management
(CKRM) \cite{ckrm2004} support more complete virtualization and should
be relatively straight-forward to integrate with Tycoon.

Another area of future work is to develop a scalable banking
infrastructure. One possibility is to physically distribute the bank
without administratively distributing it. The bank would consist of
several servers with independent account databases. A user has
accounts on some subset of the servers. A user's balance is split into
separate balances on each server. To make a transfer, users find a
server where both the payer and payee have an account and that
contains enough funds. The transfer proceeds as with a centralized
bank. Users should periodically redistribute their funds among the
servers to ensure that one server failure will not prevent all
payment.





\section{Summary}
\label{sec:conclusion}

An economic mechanism is vital for large-scale resource allocation. In
this paper, we propose a distributed market where auctioneers only
manage local resources. A user's agent sends separate bids to these
auctioneers, where each bid is for a single type of resource at that
host. The bids are continuous bids in that they stay in effect until
the user's local balance is depeleted. Resources are allocated to
users in proportion to their bids using a best-effort model. Agents
are responsible for optimizing their users' utility. 

Using our prototype implementation, we show: 1) continuous bids
reduce the burden on users by allowing them to run without frequent
interactive bidding while still making an efficient and low-latency
allocation; 2) distributed auctioneers result in very low overhead
for allocation; and 3) the best-response algorithm can optimize across
multiple markets.

\section{Acknowledgements}
\label{sec:Acknowledgements}

Several people provided key contributions without which this research
would not have been possible. David Connell assembled, installed, and
administered the Tycoon cluster. Peter Toft provided the machines in
Bristol and John Henriksson maintained them. John Janakiraman provided
several machines in Palo Alto. The economic mechanisms described here
benefited from several discussions with Leslie Fine.



\small{
\bibliographystyle{acm}
\bibliography{bibliographies/resource_allocation,bibliographies/economics,bibliographies/networking,bibliographies/overlay,bibliographies/network_performance,bibliographies/peer-to-peer,bibliographies/security,bibliographies/reputation,bibliographies/network_architecture,bibliographies/grid,bibliographies/scheduling,bibliographies/virtualization,bibliographies/algorithms}
}

\end{sloppypar}

\end{document}